\documentclass[preprint,12pt,authoryear]{elsarticle}

\usepackage{amssymb}
\usepackage{amsmath}
\usepackage{bm}
\usepackage{graphicx}
\usepackage{float}
\usepackage[table,xcdraw]{xcolor}
\usepackage{caption}
\usepackage{subcaption}
\usepackage{tikz}
\usetikzlibrary{decorations, positioning, arrows.meta, shapes.geometric}
\usepackage{pgfplots}
\pgfplotsset{compat=newest}
\usepackage{transparent}
\usepackage[colorlinks=true, allcolors=blue]{hyperref}
\usepackage{cleveref}
\usepackage{natbib}
\usepackage{multirow} 
\usepackage{dirtytalk}

\definecolor{darkred}{RGB}{200, 0, 0}

\newcommand{\OSWL}{\text{OSWL}} 
\newcommand{\OSPL}{\text{OSPL}} 
\journal{}

\begin{document}

\begin{frontmatter}



\title{Mitigating Underwater Noise from Offshore Wind Turbines via Individual Pitch Control}



\author[inst1]{Martín de Frutos}
\cortext[cor1]{Corresponding author}
\ead{m.defrutos@upm.es}
\affiliation[inst1]{organization={ETSIAE-UPM - School of Aeronautics, Universidad Politécnica de Madrid},addressline={Plaza Cardenal Cisneros 3}, 
city={Madrid},postcode={ E-28040}, country={Spain}}
\author[inst2]{Laura Botero-Bolívar}
\author[inst1,inst3]{Esteban Ferrer}

 \affiliation[inst2]{organization={Mechanical Engineering department, Universidad Industrial de Santander},
             addressline={Carrera 27 Calle 9, Bucaramanga}, 
             city={Bucaramanga},
             postcode={680001},
             country={Colombia}}

 \affiliation[inst3]{organization={Center for Computational Simulation, Universidad Politécnica de Madrid},
             addressline={Campus de Boadilla del Monte}, 
             city={Madrid},
             postcode={E-28660},
             country={Spain}}

\begin{abstract}
This paper proposes a pitch control strategy to mitigate the underwater acoustic footprint of offshore wind turbines, a measure that will soon become necessary to minimize impacts on marine life, which rely on sound for communication, navigation, and survival.

First, we quantify the underwater acoustic signature of blade‐generated aerodynamic noise from three reference turbines—the NREL 5~MW, DTU 10~MW, and IEA 22~MW—using coupling blade element momentum and coupled air–water acoustic propagation modeling. Second, we propose and implement an open‑loop individual pitch control (IPC) strategy that modulates the pitch of the blade at the blade passing frequency to attenuate the overall sound pressure level (OSPL) and the amplitude modulation (AM) of the transmitted noise. Third, we benchmark IPC performance against conventional pitch schemes. The results indicate that up to 5~dB reductions in OSPL and a decrease in AM depth 20\% can be achieved with a pitch variation $\Delta\theta \approx 5^\circ$, with small losses (5-10\%) in energy capture. These findings highlight a previously underappreciated noise pathway and demonstrate that targeted blade‐pitch modulation can mitigate its impact.
\end{abstract}




\end{frontmatter}
\section{Introduction}

Offshore wind energy is rapidly expanding as a cornerstone of the global transition to low‑carbon power, with installed capacity projected to exceed 250~GW by 2040 \citep{europeancommission2023offshoreport}. Although the underwater soundscape of offshore wind farms has traditionally focused on the noise of the foundation installation and the emissions of mechanical machinery \citep{tougaard2020loud}, an important and underappreciated path remains: the penetration of aerodynamic blade noise from the air into the marine environment. Unlike previously studied offshore noise sources, wind turbine noise persists throughout the life cycle of the wind farm. Aerodynamic noise, generated by blade–airflow interactions, has been extensively characterized in terms of its far‑field aerial propagation. However, when these sound waves impinge on the sea surface, a fraction of their energy can be transmitted  underwater \citep{Chapman1990}. This anthropogenic underwater noise has the potential to cause a masking problem, interfering with the communication, navigation, and foraging behaviors of marine fauna \citep{erbe2016communication}.

Regarding wind turbine noise mitigation techniques, most of them are implemented in the design stage \citep{deshmukh2019wind}. However, little research is devoted to wind control strategies designed to account for noise emissions. These silent control strategies could be applied to existing facilities. In our previous work, \citep{frutos2025enhancing} we introduced reinforcement learning-based wind turbine control to balance power output and noise reduction in onshore environments. In this work, we will leverage individual pitch control (IPC) to specifically reduce the aerodynamic wind turbine noise that will penetrate the air-water interface.    

Individual pitch control came about as a method that focuses mainly on reducing blade loads \citep{bossanyi2003individual}. However, with time, different IPC strategies have been designed for different applications \citep{jiang2016review}. For example, reducing power fluctuations due to tower shadowing \citep{zhang2012individual} or wake manipulation as the recent Helix approaches \citep{frederik2020helix, taschner2023performance, mohammadi2025assessment}. Regarding noise emissions, \cite{mackowski2021wind} proposed an IPC scheme to reduce the depth of characteristic modulation of amplitude of the near-field wind turbine noise. From a control perspective, load reduction strategies generally implement closed-loop conventional control methods for IPC, although recent reinforcement learning approaches have been proposed, \citep{coquelet2022reinforcement}. However, other approaches, such as Helix control, tend to design open-loop analytical IPC schemes.



In this study, we quantify the underwater acoustic footprint attributable to aerodynamic sources of the blades of large offshore horizontal-axis wind turbines. We propose an open-loop IPC strategy associated with the blade passing frequency designed to reduce underwater noise. The proposed IPC approach is evaluated on three reference wind turbines of increasing capacity: the NREL 5~MW, DTU 10~MW, and IEA 22~MW. To show the validity and generalization for turbines of increasing size, three metrics are addressed: power generation, overall sound pressure level (OSPL) and normal amplitude modulation (AM) depth. Amplitude modulation is recognized as a significant contributor to noise annoyance in humans \citep{lee2011annoyance}. 

The paper is organized as follows. In \Cref{sec:meth} a metric is defined to quantify the underwater acoustic footprint and the IPC scheme. Then, in \Cref{sec:results}, this IPC is validated for the three different wind turbines, considering its effect on different species of marine mammals. Finally, in \Cref{sec:conclusions} we come to conclusions and perspectives.


\section{Methodology}
\label{sec:meth}
An IPC control strategy to reduce the underwater acoustic footprint of wind turbines with minimal impact on power performance is presented in this section. First, how the wind turbine is modeled in both power and noise prediction is described in \Cref{sec:wt_modeling}. Next, in \Cref{sec:oswl} a metric is designed to assess the underwater acoustic footprint. Finally, the motivation for employing IPC is discussed in \Cref{sec:motivation}, and the IPC scheme is detailed in \Cref{sec:pitch_law}. 

\subsection{Wind Turbine Modeling}
\label{sec:wt_modeling}
To compute the power output and noise generation of the wind turbine, OpenFAST \citep{OpenFAST} is used. OpenFAST is an open source wind turbine simulation tool based on blade element momentum theory with coupled multi-physics modules, including aeroacoustics. The aeroacoustic model employed is the semi-empirical Brook Pope and Marcolini model \citep{brooks1989airfoil}, BPM. 
The two more significant noise sources modeled by the BPM model are the trailing edge (TE) noise mechanism and the leading edge inflow turbulence noise. The latter depends only on the inflow conditions and the rotor speed, so it cannot be controlled using an IPC strategy. In contrast, TE depends on the parameters of the boundary layer of the blade section, making it sensible to operational conditions such as the angle of the pitch of the blade. Moreover, TE noise is widely recognized as the dominant aerodynamic noise source in wind turbines, \citep{oerlemans2011wind}. Furthermore, it is especially relevant on offshore sites, where the turbulence intensity levels of the inflow wind are lower than onshore. Therefore, for this study, we only consider the trailing edge noise mechanism. 

The wind turbine controller is implemented from the DRC reference open source baseline controller \citep{mulders2018delft}. It has been modified to include an IPC controller that follows the pitch law explained in \Cref{sec:pitch_law}. 

\subsection{Wind turbine noise radiated underwater}
\label{sec:oswl}
In this section, we aim to establish a quantitative metric that characterizes the extent to which wind turbine noise can propagate into the underwater environment. Plane wave theory is used to estimate the air-water propagation of acoustic waves. Given the significant contrast in sound speed between air ($c_a$) and water ($c_w$), characterized by a speed ratio of  
$n = \frac{c_w}{c_a} \approx 4.37$,  
the transmission of airborne noise into water is constrained by Snell's law, given by:
\begin{equation}
    c_w\sin\phi = c_a\sin\alpha, 
\end{equation}
where $\alpha$ and $\phi$ are the incident and refracted angles, respectively.\\[3mm]

\noindent\textbf{Snell Cone}

Snell law explains how the acoustic waves are refracted when surpassing the air-water interface. Considering a plane interface and the air-water index of refraction $n$, only the acoustic energy radiated within a conical region, defined by a semi-angle of $\phi_{\text{lim}} = \arcsin(n^{-1}) \approx 13^\circ$, is effectively transmitted to water \citep{Chapman1990}, and we denote that region as the \say{Snell Cone}. 

\begin{figure}[htbp]
    \centering
    \begin{subfigure}[b]{0.45\textwidth}
        \centering
        \includegraphics[width=\textwidth]{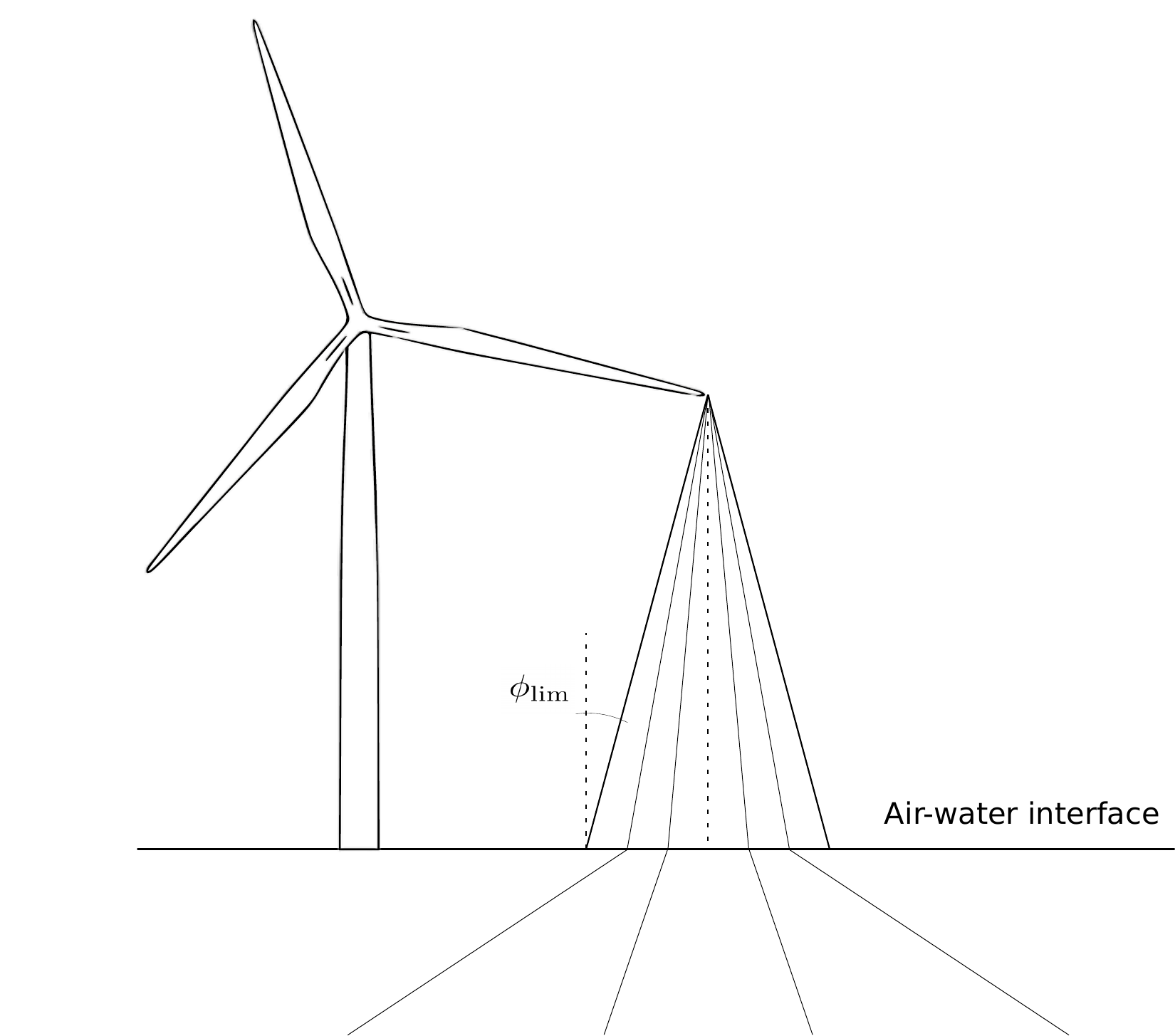}
        \caption{Wind turbine and Snell Cone illustration for one  blade. The limit angle $\phi_\mathrm{lim}$ and some sound rays following Snell's law are shown.}      \label{fig:SnellDiagram}
    \end{subfigure}
    \hfill
    \begin{subfigure}[b]{0.45\textwidth}
        \centering
        \includegraphics[width=\textwidth]{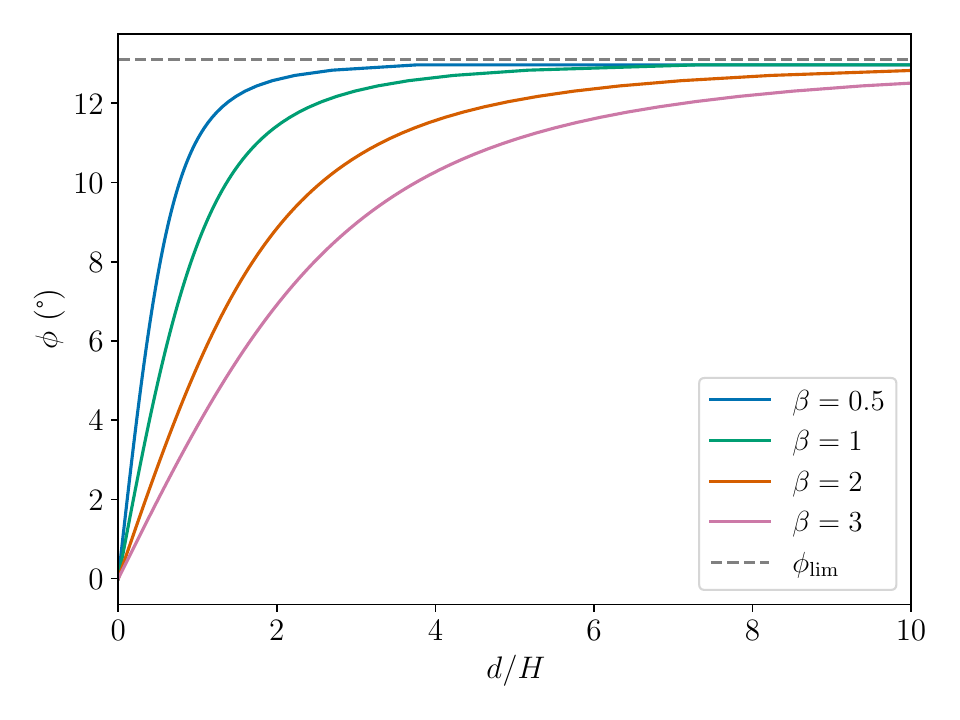}
        \caption{Snell incidence angle with distance $d$ for different depths. The sound source is located at $H$ meters above the air-water interface and the observer is located $\beta H$ meters deep.}
        \label{fig:SnellPhi}
    \end{subfigure}
    \caption{Illustration of air–water sound refraction effects.
    On the left, refraction of acoustic rays from a blade-tip noise source governed by Snell’s law.
    On the right, the effect of distance on the Snell incidence angle $\phi$.
    }
    \label{fig:Snell}
\end{figure}

Regarding wind turbine noise, most of the sound is generated at the tip of the blade~\citep{oerlemans2007location}. Therefore, we consider one Snell Cone per blade located at 95\% of the blade length. Each blade's noise is propagated underwater only through its respective Snell Cone. \Cref{fig:SnellDiagram} illustrates the Snell Cone generated by one wind turbine blade.

Additionally, as a consequence of Snell's law refraction, the sound rays that are closer to the limit angle reach much farther in the ocean. In other words, underwater observers far away from the noise source are reached by almost the limit-angle sound rays. \Cref{fig:SnellPhi} shows how the angle of incidence $\phi$ approaches the limit angle $\phi_{\mathrm{lim}}$ as the observer is further away from the noise source. Due to strong refraction at the air–water interface, distant observers primarily receive sound rays near the limiting angle. This property will later be used to estimate the overall underwater sound pressure level.

\noindent\textbf{Overall Sound Power Level that penetrates the interface}

To quantify the propagation of underwater noise, we consider how much noise irradiated by the wind turbine can exceed the interface. We can then compute the Overall Sound Power Level, OSWL, of the WT taking into account this Snell-cone propagation. We define the OSWL irradiated from the wind turbine that penetrates the air-water interface as follows: 
\begin{equation}
\label{eq:OSWL_Snell_exact}
    \OSWL(t) = 10\log_{10}\left(\frac 1{p_{\text{ref}}^2} \int_{\mathcal S}\sum_{b=1}^B S_{pp}^b(\bar x,t)\chi^b(\bar x,t)d\mathcal S\right), 
\end{equation}
where $\mathcal S$ denotes the area of the air-water interface, $S_{pp}^b(\bar x,t)$ is the total noise produced by the blade $b$ at the observer location $\bar x$ and at instant $t$, calculated in $\mathrm{Pa}^2$ and integrated over all frequencies. The function $\chi^b(\bar x,t)$ acts as a mask for the Snell Cone of each blade, whose area is $\mathcal A_{\text{SC}}^b(t)$. So, at each instant and for each observer, we only consider the noise generated by the blades whose Snell cone contains the observer. This masking function is defined as follows. 
\begin{equation}
    \label{eq:mask_snell}
    \chi^b(\bar x,t) = 
    \begin{cases}
        1, & \bar x\in\mathcal A_{\text{SC}}^b(t) \\
        0, & \bar x\notin\mathcal A_{\text{SC}}^b(t).
    \end{cases}
\end{equation}

\begin{figure}[htbp]
    \centering
    \includegraphics[width=0.7\linewidth]{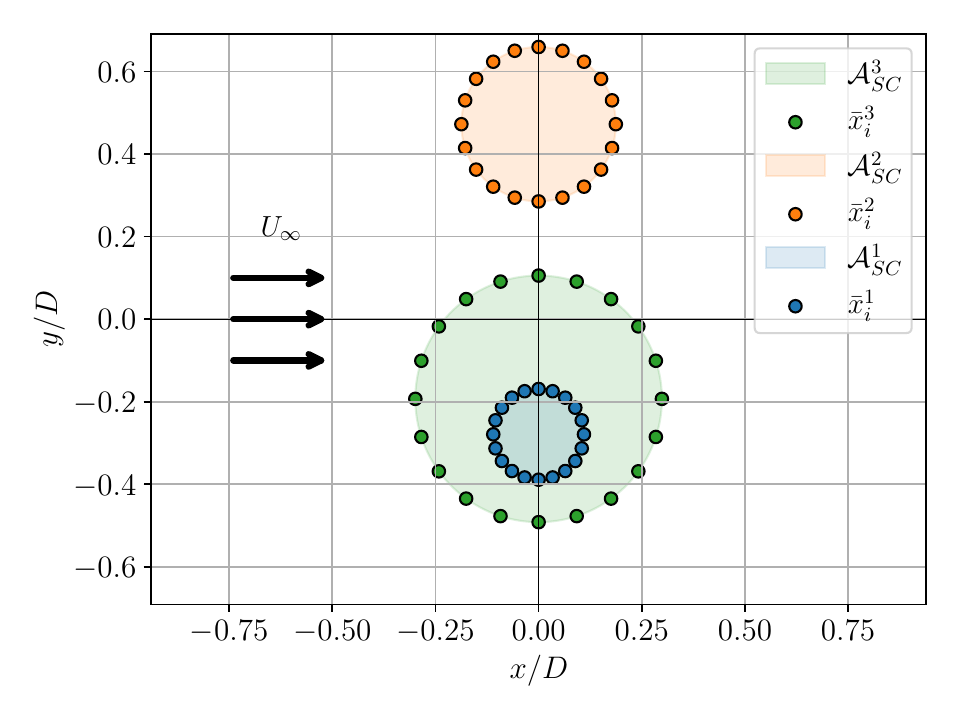}
    \caption{Observers configuration employed to estimate $\overline\OSPL$ at certain time instant for each wind turbine blade. $N_c = 20$ observation points are uniformly distributed along the Snell Cone border. The wind direction is from $-x$ to $+x$, as indicated, and the vertical axis corresponds to $z$. The view is from above the rotor plane.}
    \label{fig:ObserverMap}
\end{figure}

In practice, we use a BPM method to predict wind turbine noise. This semi-empirical methods require to be evaluated at some specific observer locations. Therefore, we need to discretize \cref{eq:OSWL_Snell_exact}. 
At each instant of time, we compute the noise of the wind turbine in the observers $N_c$ at the intersection between each Snell Cone blade and the air-water interface; those coordinates are denominated as $\bar x_i^b(t)$, with $i$ from 1 to $N_c$. We only consider observers located at the limiting angle, near the intersection of the Snell cone with the sea surface, since these positions are the most relevant for far-field underwater acoustics (see \cref{fig:SnellPhi}). \Cref{fig:ObserverMap} illustrates the observers used to estimate the overall sound power level integral. From the noise at the selected observer locations, we can compute $\overline\OSPL(t)$, defined as the overall sound pressure level averaged on the blade's Snell Cone regions at each time instant,

\begin{equation}
\label{eq:OSWL_Snell_approx}
    \overline\OSPL(t) \approx 10\log_{10}\left(\frac1{p_{\text{ref}}^2}\sum_{b=1}^B \left(\frac1{N_c}\sum_{i=1}^{N_c}S_{pp}^b(\bar x_i^b(t),t)\right)\frac{\mathcal A^b_{\text{SC}}(t)}{\mathcal A_{\text{SC}}(t)}\right),
\end{equation}
where $\mathcal A_{\text{SC}}(t)$ is the sum of the three areas of the Snell cone. The $\mathcal A^b_{\text{SC}}(t)/\mathcal A_{\text{SC}}(t)$ ratio serves to weight each blade contribution, as the number of observers is the same regardless of the Snell Cone area.

Notice that $\overline{\OSPL}$ represents the overall pressure level that would be heard on average in the Snell cone region of the three blades. However, this sound pressure level is above the interface. We consider the region where the noise will propagate, but we do not employ any transmission loss modeling. To later estimate the actual sound pressure level at an underwater receiver, the present framework would need to be coupled with an acoustic propagation model. A general formulation for such an estimate is given by:
\begin{equation}
\label{eq:propagation_losses}
    \mathrm{SPL}(f) = \widehat{\mathrm{SPL}}(f)+\Delta L - 10\log_{10}(4\pi d^2) - \alpha_w(f) d, 
\end{equation} 
where $\mathrm{SPL}(f)$, is the sound pressure level spectrum at a certain underwater receiver, $\widehat{\mathrm{SPL}}(f)$ is the sound pressure level spectrum averaged on the Snell Cone and over a rotation, $\alpha_wd$ accounts for frequency-dependent acoustic attenuation in water, and $\Delta L$ considers the transmission loss associated with the air-water interface and any other effect that is not captured in the spherical spreading, $10\log_{10}(4\pi d^2)$. 

Although not directly modeling underwater propagation, $\overline\OSPL$ captures the phenomena that will be used in the proposed IPC strategy, that is, the near-field directivity of wind turbine noise and the influence of blade pitch on the generation of aerodynamic noise. Thus, it is a suitable metric for validating the efficacy of the IPC. 

\subsection{Why Individual Pitch Control?}
\label{sec:motivation}

\begin{figure}[htbp]
    \centering
    \begin{subfigure}[b]{0.45\textwidth}
        \centering
        \includegraphics[width=\linewidth]{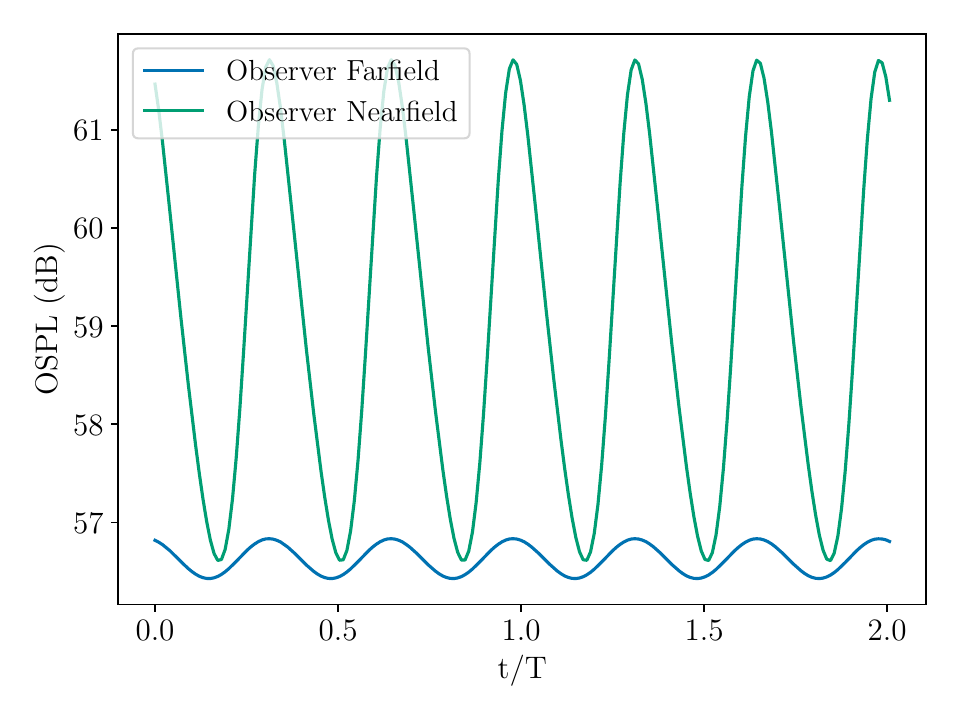}
        \caption{OSPL directivity effect for near-field and far-field observers. The observers are located 50 m downstream and 200 m downstream respectively.}
        \label{fig:ObserversDirectivity}
    \end{subfigure}
    \hfill
    \begin{subfigure}[b]{0.45\textwidth}
        \centering
        \includegraphics[width=\linewidth]{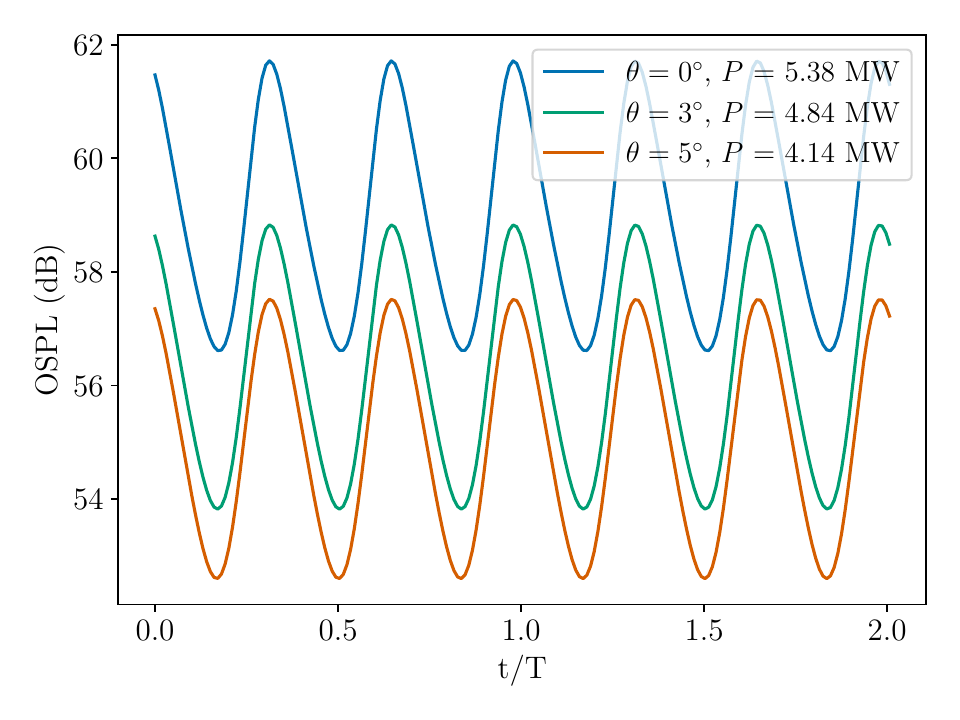}
        \caption{Power, $P$ and noise sensitivity with pitch blade angle, $\theta$ for an observer in the near-field, 50 m downstream of the wind turbine.}
        \label{fig:PitchSensitivity}
    \end{subfigure}
    \caption{Wind turbine noise behavior for different observer positions and blade pitch angle values over two rotor revolutions. Results presented for the NREL 5~MW wind turbine.}
    \label{fig:ipc_motivation}
\end{figure}

There are two different effects that can be leveraged by using IPC to reduce underwater noise propagation. First, due to the characteristics of air-water acoustic transmission, underwater noise perception is primarily influenced by the near-field region, as wind turbine blade noise propagates only within its respective Snell Cone, located beneath the turbine. In this region, the noise fluctuations are higher due to the directivity of the source, as illustrated in \Cref{fig:ObserversDirectivity}. There, the noise generation is dominated by the blade in the downward position, as reported by \cite{oerlemans2007location}.
Second, the blade pitch angle is known to significantly affect both aerodynamic power output and noise generation. Increasing the pitch angle from nominal values typically leads to reductions in both power and noise levels; see \cite{maizi2017reducing,frutos2025enhancing}. However, in the near-field region, noise oscillations can overlap for different pitch angle values, as shown in \Cref{fig:PitchSensitivity}. Based on this observation, a potential strategy is to design an IPC scheme that increases the pitch angle exclusively for the downward-moving blade. This approach would reduce the overall noise by aligning it more closely with low-noise pitch settings, while maintaining high power output, as only one blade is modified at a time.
\subsection{Pitch Law}
\label{sec:pitch_law}
We propose an IPC scheme that pitches the blade in the downward position. An analytical law is proposed to ensure smooth pitch variation, which defines each blade pitch angle $\theta^b$ depending on the local blade azimuth angle $\Psi^b$. 
\begin{equation}
    \label{eq:pith_law}
    \theta^b(\Psi^b) =
    \begin{cases} 
        \theta_1 + \frac12(\theta_2 - \theta_1)(1 + \tanh((\Psi^b - \Psi_1) k)), & \text{if } \Psi^b \leq \Psi_c \\[10pt]
        \theta_2 + \frac12(\theta_1 - \theta_2)(1 + \tanh((\Psi^b - \Psi_2) k)), & \text{if } \Psi^b > \Psi_c. 
    \end{cases}
\end{equation}

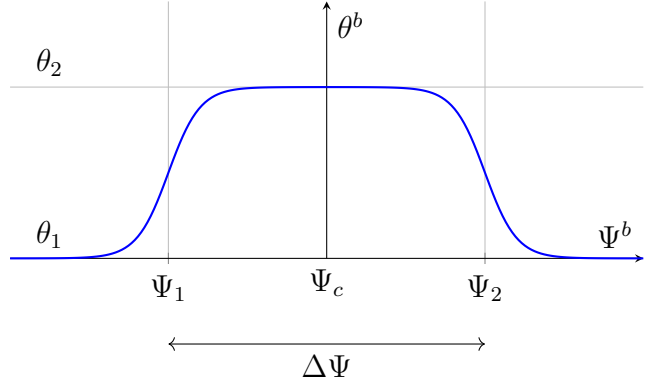
\begin{figure}[htbp]
    \centering
    \begin{tikzpicture}
    \begin{axis}[
        axis lines = middle,
        xlabel = {$\Psi^b$},
        ylabel = {$\theta^b$},
        xmin = -2, xmax = 2,
        ymin = 0, ymax = 1.5,
        samples = 100,
        domain = -2:2,
        grid = major,
        width=10cm,
        height=5cm,
        xtick={-1,1},
        ytick={0,1},
        xticklabels={$\Psi_1$,$\Psi_2$}, 
        yticklabels={}, 
        clip=false,
    ]
        
        \addplot[blue, thick, domain=-2:0] {0.5+0.5*tanh((x+1)*5)};
        \addplot[blue, thick, domain=0:2] {0.5-0.5*tanh((x-1)*5)};
        \draw[<->] (-1,-0.5) -- (1,-0.5) node[midway, below]{$\Delta\Psi$};
        \node[below] at (0,0) {$\Psi_c$};
        \node[above] at (-1.75,0) {$\theta_1$};
        \node[above] at (-1.75,1) {$\theta_2$};
    \end{axis}

\end{tikzpicture}
    \caption{Pitch law for each blade depending on its phase.}
    \label{fig:pitch_law}
\end{figure}

This law sets a high-noise/power pitch value $\theta_1$ and alternates to a low noise-power pitch value $\theta_2$ between the local azimuth positions $\Psi_1$ and $\Psi_2$. We define the central phase azimuth position $\Psi_c=\frac12(\Psi_2+\Psi_1)$ and the phase width $\Delta\Psi=\Psi_2-\Psi_1$. The transition between the values is made using a hyperbolic tangent function with parameter $k$. The local origin of the azimuth position $\Psi^b=0$ is when the blade $b$ is vertically facing upward. \Cref{fig:pitch_law} illustrates this analytical pitch law. 

The parameters of \cref{eq:pith_law} determine the power-noise trade-off of the IPC scheme. We can give some bounds to $k$ and $\Delta\Psi$ to ensure that the pitch law is applicable to existing offshore wind turbines that can effectively alternate between pitch values. 

Usually, wind turbine manufacturers set a maximum pitch rate. For example, in the NREL 5~MW turbine~\citep{NRELdefinition}, the maximum pitch rate allowed in the control is $10^\circ$/s. This constraint can be used to bound the maximum steepness of the hyperbolic tangent. Hence, a maximum value for $k$ can be obtained, 
\begin{equation}
\label{eq:k_max_cond}
    \max\frac{d\theta^b}{dt} = \Omega\left.\frac{d\theta^b}{d\Psi}\right\rvert_{\Psi_1} = \Omega k\Delta\theta \le \dot\theta_{\max} \Rightarrow k_{\max} = \frac{\dot\theta_{\max}}{\Omega\Delta\theta}, 
\end{equation}
where $\Delta\theta=\theta_2-\theta_1$ is the pitch jump and $\Omega$ is the speed of the wind turbine rotor. Similarly, we can obtain the minimum phase width required to ensure that the pitch reaches $\theta_2$. Since the transition is a hyperbolic tangent, the final pitch value is reached only asymptotically. We can define the minimum phase to reach at least 95\% of $\theta_2$,   
\begin{equation}
\label{eq:psi_min_cond}
    \tanh\left(\frac{\Delta\Psi k}{2}\right)\approx 1 \Rightarrow \Delta\Psi_{\min} = \frac{2\Omega\Delta\theta}{\dot\theta_{\max}}\operatorname{atanh}(0.95). 
\end{equation}
\section{Results and discussion}
\label{sec:results}
The IPC scheme proposed in \Cref{sec:pitch_law} is evaluated for three offshore wind turbines: the NREL 5~MW \citep{NRELdefinition}, the DTU 10~MW \citep{DTUdefinition}, and the IEA 22~MW\citep{IEAdefinition}. These wind turbines span a wide range of geometric operational conditions; see \Cref{tab:WT_nominal_cond}. All results presented are obtained using nominal wind speed and rotational speed conditions. 

In \Cref{sec:power-noise}, the power-noise trade-off for these three offshore wind turbines is discussed. Two different IPC schemes are proposed and tested. Finally, in \Cref{sec:spec_animals}, we quantify the effectiveness of the IPC strategy in different groups of marine animals.

\begin{table}[htbp]
\begin{tabular}{@{}lccc@{}}
\hline
Characteristic & NREL~5~MW  & DTU~10~MW & IEA~22~MW\\
\hline
Hub height [m] & 90.0   & 119.0 & 170.0 \\
Rotor diameter [m] & 126.0   & 178.4  & 284.0  \\
Nominal wind velocity [m/s]    & 11.4   & 11.4  & 11.13  \\
Rotor angular velocity [rpm]    & 12.1   & 9.6  & 7.1  \\
Blade tip velocity [m/s] & 79.0  & 90.0  & 102.0  \\
Blade Pitch [deg.] & 0.0   & 0.0  & 4.12 \\
\hline
\end{tabular}
\caption{Geometrical and operational (rated) conditions of three large offshore wind turbines.} \label{tab:WT_nominal_cond}
\end{table}

\subsection{Power and noise balance for three offshore wind turbines}
\label{sec:power-noise}
The trade-off between power extraction and wind turbine noise is primarily governed by the pitch angles defined by the pitch control law (see \cref{eq:pith_law}), specifically $\theta_1$ and $\theta_2$. The selection of these values reflects the relative importance assigned to power generation versus noise reduction.

The high-noise/power angle $\theta_1$ corresponds to the nominal setting of the pitch angle. This is typically optimized for maximum power output. From a control perspective, $\theta_1$ would be the collective pitch angle imposed by a conventional variable-speed controller. In contrast, $\theta_2$ serves as a tunable parameter to balance the power-noise trade-off. Increasing the pitch angle beyond nominal conditions generally results in a simultaneous reduction in both power and noise levels.

This section presents a brief discussion on the impact of pitch jump $\Delta\theta=\theta_2-\theta_1$ on power generation and noise emissions. Two different IPC schemes are studied, with pitch jumps of $\Delta\theta^1=3^\circ$ and $\Delta\theta^2=5^\circ$. A higher value of the pitch jump requires a larger phase width, so the pitch has enough time to transition. Hence, different phase widths are used, $\Delta\Psi^1=120^\circ$ and $\Delta\Psi^2=150^\circ$. Notice that both are larger than $\Delta\Psi_{\text{min}}$ defined by \cref{eq:psi_min_cond} for each case. The value of both pitch laws $k$ is set as the maximum allowed pitch rate; see \cref{eq:k_max_cond}. The central phase selected for both strategies is $\Psi_c=135^\circ$, which corresponds to the middle azimuth position in the downward quarter.

\begin{figure}[htbp]
    \centering
    \begin{subfigure}[b]{0.45\textwidth}
        \centering
        \includegraphics[width=\textwidth]{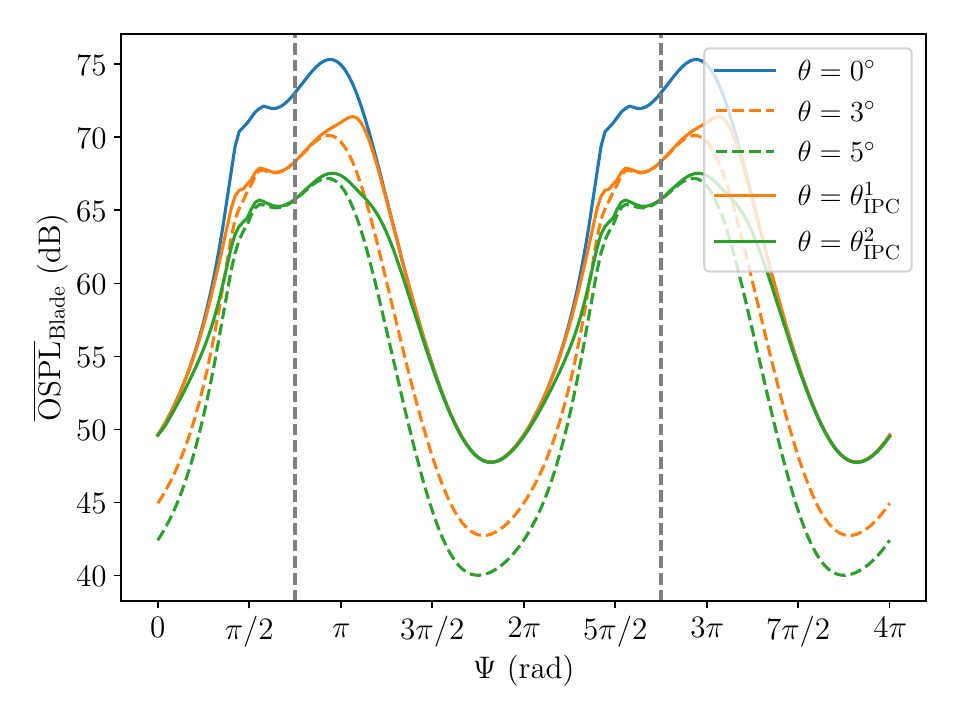}
        \caption{Blade $\overline\OSPL$. The dashed lines denotes the central phase $\Psi_c$ of \cref{eq:pith_law} pitch law.}
        \label{fig:DTUBlade_pitch}
    \end{subfigure}
    \hfill
    \begin{subfigure}[b]{0.45\textwidth}
        \centering
        \includegraphics[width=\textwidth]{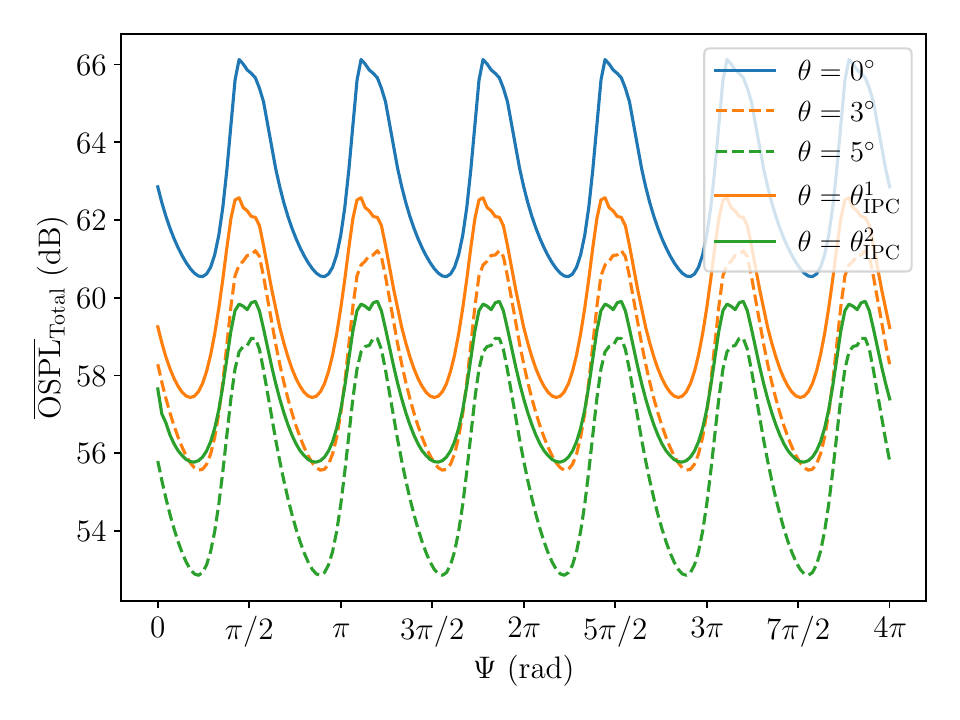}
        \caption{Total $\overline\OSPL$. $\Psi$ denotes the azimuth phase of the first blade. }
        \label{fig:DTUTotal_pitch}
    \end{subfigure}
    \caption{Comparison of different pitch strategies on the $\overline\OSPL$ for the DTU 10 MW wind turbine for two rotor revolutions.}
    \label{fig:DTU_IPC_pitch}
\end{figure}

For illustration, in \Cref{fig:DTU_IPC_pitch} we present the average overall sound pressure level on the Snell Cone ($\overline{\mathrm{OSPL}}$) produced by each IPC strategy applied to the DTU 10~MW wind turbine. The results corresponding to constant pitch values—namely the nominal angle and increments of $+3^\circ$ and $+5^\circ$—are also included. \Cref{fig:DTUBlade_pitch} illustrates how the pitch law transitions from following the low noise $\overline\OSPL$ curve when the blade is oriented downward to aligning with the high noise curve when the blade noise radiated in the Snell Cone region is minimal. This showcases the pitch law's ability to reduce noise by targeting the most acoustically sensitive positions. Finally, the result $\overline{\mathrm{OSPL}}$ for the entire wind turbine is shown in \cref{fig:DTUTotal_pitch}. There are some phenomena to consider when analyzing the noise from the three blades. First, it is important to note that the lowest noise levels observed in the single-blade case (see \cref{fig:DTUBlade_pitch}) are not fully recovered in the full rotor configuration (see \cref{fig:DTUTotal_pitch}). This is due to the $2\pi/3$ phase difference between blades, when one blade is in a low-noise position, the other two occupy different angular positions. As a result, total wind turbine noise exhibits an overlap of high- and low-noise regions, leading to a reduction in the depth of modulation of the amplitude of the signal, $\overline{\mathrm{OSPL}}(t)$. Another important remark is that the maximum value of $\overline{\mathrm{OSPL}}(t)$ for the entire rotor is lower than the maximum noise level observed for a single blade. This occurs because $\overline{\mathrm{OSPL}}(t)$ represents the average sound pressure level over the Snell Cone region. In the case of three blades, the averaging is performed over the union of the three Snell Cones, which covers a larger area. Since not all blades emit maximum noise simultaneously, spatial averaging across this extended region results in a lower overall peak level.

To evaluate the effectiveness of the two IPC schemes, three key performance metrics are compared across the three wind turbines:
\begin{enumerate}
    \item Power loss with respect to the nominal operating condition. $$\mathrm{Power~Loss} = \frac{\mathrm{Power}}{\mathrm{Power~nominal}}\cdot 100\%.$$
    \item Noise generation. We denote $\widehat\OSPL$ to the overall sound pressure level averaged over the Snell Cone area and over one rotation. Notice that it is either integrating all frequencies of $\widehat{\mathrm{SPL}}(f)$ or time averaging $\overline\OSPL(t)$, both previously introduced.
    \item Normal amplitude modulation depth (AM) of $\overline\OSPL(t)$. Normal AM measures the amplitude of the overall sound pressure level oscillations caused by the directivity of the dominant trailing edge noise source combined with the time-varying position and orientation of the rotating blades. To compute the AM depth, the method proposed by \cite{lee2011annoyance} is followed. 
\end{enumerate}

The results are summarized in \Cref{tab:IPC_pitch}.
As expected, the low-pitch jump IPC scheme, denoted $\theta_\mathrm{IPC}^1$, yields the smallest power losses relative to the nominal case. However, this scheme achieves a more modest reduction in $\widehat{\mathrm{OSPL}}$ compared to the higher pitch jump variant, $\theta_\mathrm{IPC}^2$. In particular, both IPC strategies demonstrate the ability to reduce AM with respect to all constant pitch angle configurations. 

The choice between $\theta_\mathrm{IPC}^1$ and $\theta_\mathrm{IPC}^2$ ultimately depends on the relative priority given to power production versus noise mitigation. The low pitch jump scheme $\theta_\mathrm{IPC}^1$, incurs relatively minor power losses—approximately 5\% depending on the turbine—while achieving a $\sim$3 dB reduction in $\widehat{\mathrm{OSPL}}$. In contrast, the higher pitch jump scheme ($\theta_\mathrm{IPC}^2$) can reduce noise by up to 5 dB, but at the cost of power losses reaches approximately 10\%.

\begin{table}[htbp]
    \centering
    \begin{tabular}{l|l|ccccc}
    \hline
    \textbf{WT} & \textbf{Metric} & $\theta_{\text{nom}}$ & $3^\circ$ & $5^\circ$ & $\theta_\mathrm{IPC}^1$ & $\theta_\mathrm{IPC}^2$ \\
    \hline
    \multirow{3}{*}{NREL} & Power loss (\%) & 0.00 & 10.16 & 23.05 & 3.19 & 8.45 \\
     & $\widehat\OSPL$ (dB) & 55.97 & 53.68 & 52.82 & 54.36 & 53.45 \\
     & AM (dB) & 7.58 & 7.79 & 8.02 & 7.30 & 7.22 \\
    \hline
    \multirow{3}{*}{DTU} & Power loss (\%) & 0.00 & 4.89 & 15.39 & 1.36 & 5.21 \\
     & $\widehat\OSPL$ (dB) & 63.39 & 58.66 & 56.25 & 60.14 & 57.90 \\
     & AM (dB) & 6.42 & 6.62 & 7.10 & 5.97 & 5.00 \\
    \hline
    \multirow{3}{*}{IEA} & Power loss (\%) & 0.00 & 18.62 & 36.65 & 6.17 & 14.73 \\
     & $\widehat\OSPL$ (dB) & 64.34 & 59.15 & 56.96 & 60.81 & 58.58 \\
     & AM (dB) & 6.80 & 7.50 & 7.75 & 5.90 & 5.10 \\
    \hline
    \end{tabular}
    \caption{Comparison of three performance metrics for different pitch strategies across three offshore wind turbines. The evaluated pitch strategies are: the nominal pitch angle ($\theta_\mathrm{nom}$), fixed pitch increments of $3^\circ$ and $5^\circ$, and two IPC schemes designed to alternate between the nominal condition and these respective increments.}
    \label{tab:IPC_pitch}
\end{table}

\subsection{Influence on marine life}
\label{sec:spec_animals}
\begin{figure}[htbp]
    \centering
    \begin{subfigure}[b]{0.45\linewidth}
        \centering
        \includegraphics[width=\linewidth]{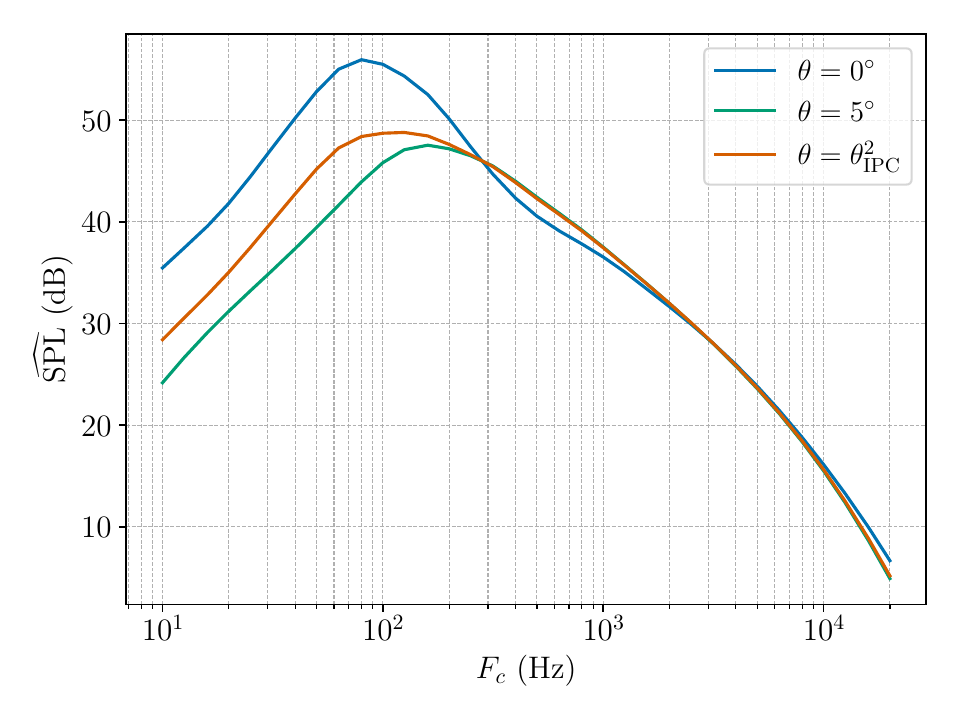}
        \caption{$\frac13$-octave $\widehat{\mathrm{SPL}}$ spectrum for the three pitch configurations for the DTU 10 MW wind turbine.}
        \label{fig:Spectrum_DTU}
    \end{subfigure}
    \hfill
    \begin{subfigure}[b]{0.45\linewidth}
    \centering
    \includegraphics[width=\linewidth]{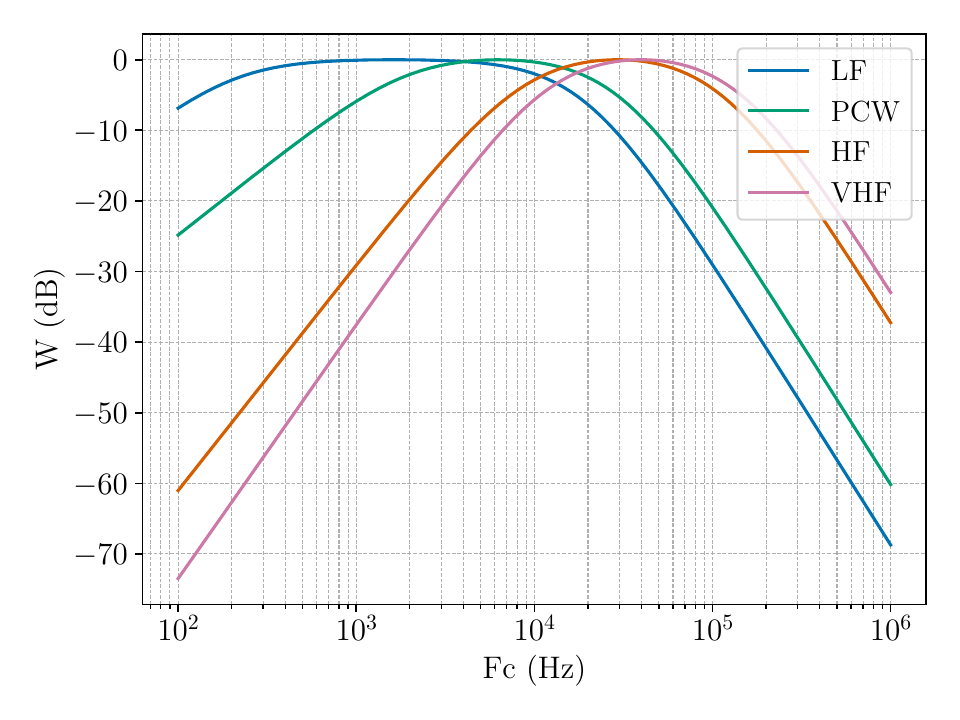}
    \caption{Frequency weighting curves proposed by \cite{national2016technical} for different marine animal groups.}
    \label{fig:FiltersAnimals}
    \end{subfigure}
    \caption{Frequency-domain analysis of wind turbine noise considering IPC and its relevance to marine fauna.}
\end{figure}

Similarly to humans, marine animals do not hear equally along all frequencies. To characterize how different species of marine mammals perceive sound, \citet{southall2019marine} grouped different species into functional hearing groups based on several experimental studies. In this study, we consider the following groups: 
\begin{itemize}
    \item \textbf{LF}-cetaceans: Mysticete whales, including minke whale.
    \item \textbf{HF}-cetaceans: Most odontocetes, including the white-beaked dolphin and the pilot whale.
    \item \textbf{VHF}-cetaceans: Narrow-band high-frequency odontocetes, including harbour porpoise
    \item \textbf{PCW}-phocid seals: True seals, including harbor seal and gray seal.
\end{itemize}
Regarding humans, the consensus is that weighting the SPL spectrum with a curve roughly resembling the inverted audiogram, the so-called dBA-weighting, provides the best overall prediction of the risk of injury. A similar procedure can be performed for the different functional hearing groups \citep{tougaard2021thresholds}. 
\cite{national2016technical} proposed analytical expressions for these marine weights (or filters) for the different functional groups; see \cref{fig:FiltersAnimals}. 

These group filters are used to assess the reduction in sound pressure level experienced by different groups of marine mammals as a result of the implementation of the IPC scheme. Using the high-pitch jump IPC strategy $\theta_\mathrm{IPC}^2$ the $\frac13$-octave $\widehat{\mathrm{SPL}}$ spectrum is shown in \Cref{fig:Spectrum_DTU} for the DTU 10 MW wind turbine. As observed, the spectrum corresponding to the IPC scheme consistently falls between the spectra obtained with fixed pitch angles, illustrating the intermediate acoustic footprint of the IPC approach.

\Cref{tab:FilterReduction} presents the overall reduction in the sound pressure level $\Delta C$ due to the IPC scheme relative to the nominal conditions considering different filters for marine mammals. Note that the unfiltered OSPL reduction scales with the size of the wind turbine, which indicates that this IPC control strategy is particularly suitable for large wind turbines, where the wind turbine noise impact is more significant.

As shown in \cref{fig:Spectrum_DTU}, the spectral response of $\widehat{\mathrm{SPL}}$ to pitch modifications mainly affects the mid to low frequency range. The IPC-induced noise reduction is concentrated within this frequency band, whereas reductions at higher frequencies are limited, and in some cases, a slight amplification may occur. This results in negative $\Delta C$ values (i.e., an increase in SPL) for hearing groups sensitive to high frequencies. The extent of this effect depends on the spectral characteristics of each wind turbine, as illustrated in \Cref{tab:FilterReduction}. However, note that in the high-frequency range where these species are most sensitive, the baseline wind turbine noise spectrum is already relatively low. Thus, even if the IPC strategy leads to minor increases in SPL at these frequencies, the absolute levels remain low and are unlikely to pose significant acoustic risk.

In general, the IPC strategy proves to be particularly effective for marine species sensitive to mid- and low-frequency noise. In particular, the LF hearing group exhibits the largest reduction, exceeding 3 dB for both the DTU and the IEA wind turbines.

\begin{table}[htbp]
    \centering
    \begin{tabular}{l | c c c c c}
    \hline
    \textbf{WT} & \textbf{Unfiltered} & \textbf{LF} & \textbf{PCW} & \textbf{HF} & \textbf{VHF} \\ 
    \hline
    NREL & 2.58 & 1.29 & -0.46 & -0.97 & -0.81 \\ 
    DTU & 5.35 & 3.13 & 0.59 & 0.09 & 0.28 \\ 
    IEA & 5.60 & 3.63 & 0.84 & -0.28 & -0.29 \\ 
    \hline
    \end{tabular}
    \caption{$\widehat\OSPL$ reduction (dB) using the proposed IPC strategy, $\theta_\mathrm{IPC}^2$, compared to nominal conditions for the three wind turbines studied and applying different marine animal filters.}
    \label{tab:FilterReduction}
\end{table}

As a final remark, note that the $\widehat\OSPL$ reductions due to the control $\Delta C$, are computed above the sea surface. These reductions are still valid at underwater receivers, as all the propagation losses associated with that position do not depend on the control strategy followed; see \cref{eq:propagation_losses}.  

\section{Conclusions}
\label{sec:conclusions}

In this work, we have quantified how much aerodynamic noise generated by large horizontal-axis offshore wind turbines can penetrate the air–sea interface and contribute measurably to the underwater acoustic environment.

To mitigate this airborne‑to‑underwater noise, we developed and evaluated an open‑loop IPC strategy. By modulating blade pitch at the blade passing frequency, this IPC approach substantially reduces overall sound pressure level and dampens amplitude modulation - a phenomenon known to cause annoyance in humans and likely to affect marine animals similarly. 
By combining the analyzes of three representative turbines (NREL 5~MW, DTU 10~MW, and IEA 22~MW), we quantified the reductions in sound pressure level due to the proposed IPC method. Furthermore, we computed its effect on marine mammals considering their hearing sensibility. 

Our method achieves noise reduction with minimal impact on power performance, confirming a variation in pitch $\Delta\theta \approx 3^\circ$ as an effective compromise between acoustic mitigation and power generation.

Furthermore, our results show that the greatest noise reduction is achieved in the low to mid-frequency range, where the wind turbine noise is most pronounced. They also indicate that pitch control may be ineffective as a mitigation strategy for marine mammals with high frequency hearing. Consequently, the real effect of the IPC strategy will depend on the specific local marine fauna. Therefore, a site-specific ecological assessment should be performed prior to implementing wind turbine noise control measures.

Looking ahead, integrating closed-loop feedback and reinforcement learning algorithms \citep{frutos2025enhancing,coquelet2022reinforcement} could further optimize pitch schedules in real time, taking into account the changing wind, sea state, and biological risk factors. By embedding aerodynamic noise considerations in environmental impact assessments and turbine control design, this study paves the way for a more harmonious coexistence between offshore wind development and marine ecosystems.

\section*{Acknowledgments}
This research has received funding from the European Union (ERC, Off-coustics, project number 101086075). Also, the authors acknowledge the support of Universidad Industrial de Santander and the Energy and Environment research group (GIEMA). Views and opinions expressed are, however, those of the authors only and do not necessarily reflect those of the European Union or the European Research Council. Neither the European Union nor the granting authority can be held responsible for them.

\bibliographystyle{unsrtnat}
\bibliography{cas-refs}
\end{document}